\def\Msol{\hbox{M$_\odot$}}

\def\kms{\hbox{km$\,$s$^{-1}$}}
\def\cmt{\hbox{cm$^{-3}$}}

\def\one{\,{\sc i}}             
\def\two{\,{\sc ii}}
\def\three{\,{\sc iii}}

\documentclass{emulateapj}

\usepackage{amsmath}
\usepackage{amssymb}
\usepackage{mathrsfs}
\usepackage[usenames,dvipsnames]{color,colortbl}
\usepackage{soul} 
\usepackage{upgreek}
\usepackage{graphicx}
\usepackage{overpic}
\usepackage{threeparttable}

\shorttitle{An optical--NIR study of three SSCs in M82}
\shortauthors{M.\ S.\ Westmoquette et al.}

\begin{document}
\defcitealias{smith06}{Paper~I}
\defcitealias{westm07c}{Paper~II}


\title{An optical--near-IR study of a triplet of super star clusters in the starburst core of M82
\altaffilmark{1}}
\author{M.\ S.\ Westmoquette\altaffilmark{2}}
\author{N.\ Bastian\altaffilmark{3,4}}
\author{L.\ J.\ Smith\altaffilmark{5}}
\author{A.\ C.\ Seth\altaffilmark{6}}
\author{J.\ S.\ Gallagher III\altaffilmark{7}}
\author{R.\ W. O'Connell\altaffilmark{8}}
\author{J.\ E.\ Ryon\altaffilmark{7}}
\author{S.\ Silich\altaffilmark{9}}
\author{Y.\ D.\ Mayya\altaffilmark{9}}
\author{C.\ Mu\~{n}oz-Tu\~{n}\'{o}n\altaffilmark{10}}
\author{D.\ Rosa Gonz\'{a}lez\altaffilmark{9}}

\altaffiltext{1}{Based on observations with the NASA/ESA \textit{Hubble Space Telescope} under program 11641 and the Gemini-North telescope under program GN-2010B-Q-4.}
\altaffiltext{2}{European Southern Observatory, Karl-Schwarzschild-Str. 2, 85748 Garching bei M\"{u}nchen, Germany (westmoquette@gmail.com)}
\altaffiltext{3}{Excellence Cluster Universe, Boltzmannstrasse 2, 85748 Garching bei M\"{u}nchen, Germany}
\altaffiltext{4}{Astrophysics Research Institute, Liverpool John Moores University, 146 Brownlow Hill, Liverpool L3 5RF, UK}
\altaffiltext{5}{Space Telescope Science Institute and European Space Agency, 3700 San Martin Drive, Baltimore, MD 21218, USA}
\altaffiltext{6}{University of Utah, Salt Lake City, UT, USA}
\altaffiltext{7}{Department of Astronomy, University of Wisconsin-Madison, 5534 Sterling, 475 North Charter St., Madison WI 53706, USA}
\altaffiltext{8}{Department of Astronomy, University of Virginia, P.O. Box 3818, Charlottesville, VA 22903, USA}
\altaffiltext{9}{Instituto Nacional de Astrof\'{i}sica, Optica y Electronica, Luis Enrique Erro 1, Tonantzintla, C.P. 72840, Puebla, Mexico}
\altaffiltext{10}{Instituto de Astrof\'{i}sica de Canarias, C/v\'{i}a L\'{a}ctea s/n, 38200, La Laguna, Tenerife, Spain}

\begin{abstract}
We present \textit{HST}/STIS optical and Gemini/NIFS near-IR IFU spectroscopy, and archival \textit{HST} imaging of the triplet of super star clusters (A1, A2 and A3) in the core of the M82 starburst. Using model fits to the STIS spectra, and the weakness of red supergiant CO absorption features (appearing at $\sim$6~Myr) in the NIFS H-band spectra, the ages of A2 and A3 are $4.5\pm1.0$~Myr.  A1 has strong CO  bands, consistent with our previously determined age of $6.4\pm0.5$~Myr. The photometric masses of the three clusters are 4--$7\times10^5$~\Msol, and their sizes are $R_{\rm eff}=159$, 104, 59~mas ($\sim$2.8, 1.8, 1.0~pc) for A1,2 and 3. The STIS spectra yielded  radial velocities of $320\pm2$, $330\pm6$, and $336\pm5$~\kms\ for A1,2, and 3, placing them at the eastern end of the $x_2$ orbits of M82's bar. Clusters A2 and A3 are in high density (800--1000~\cmt) environments, and like A1, are surrounded by compact H\two\ regions. We suggest the winds from A2 and A3 have stalled, as in A1, due to the high ISM ambient pressure. We propose that the 3 clusters were formed \textit{in-situ} on the outer $x_2$ orbits in regions of dense molecular gas subsequently ionized by the rapidly evolving starburst.  The similar radial velocities of the 3 clusters and their small projected separation of $\sim 25$~pc suggest that they may merge in the near future unless this is prevented by velocity shearing.
\end{abstract}

\keywords{galaxies: evolution -- galaxies: individual: M82 -- galaxies: ISM -- galaxies: starburst -- galaxies: star clusters}

\section{Introduction} \label{intro}

M82 is the archetype nearby \citep[3.6~Mpc, $1'' = 17.5$~pc;][]{freedman94} starburst galaxy \citep{oconnell78, oconnell95}. The current ($\sim$10~Myr) starburst activity is concentrated in a $\sim$500~pc ($\sim$30$''$) region centred on the nucleus. From \textit{Hubble Space Telescope} (\textit{HST}) imaging, the starburst is known to consist of a number of prominent, high surface-brightness clumps, first identified and labelled by \citet{oconnell78}. \citet{oconnell95} argued that these clumps represent the parts of the starburst core which are the least obscured along the line of sight. These clumps contain many young massive star clusters \citep{oconnell95, mccrady03, mayya08}, and it is presumed that the combined energy from these clusters \citep{t-t03} and the SNe known to be distributed throughout the central starburst zone \citep{fenech08} is what drives the famous H$\alpha$- and X-ray-bright superwind \citep{t-t97a, t-t98, shopbell98, ohyama02, stevens03a, engelbracht06, strickland07, westm07c, westm09a}.

M82 is a dynamically complex system, perturbed significantly by its gravitational encounter with M81 some $2\times10^8$~yrs ago \citep{yun_ho_lo93}. The consequences of this encounter have been playing out ever since. Evidence from the age-dating of a large sample of star clusters \citep{konstantopoulos09}, and of stellar populations in the nucleus \citep{forster03} and the disk \citep{davidge08}, suggests that initially a disk-wide burst of star formation took place, followed by a recent burst concentrated only in the nuclear regions \citep[e.g.][]{mayya06,beirao08}.
\citet{gall_smith99} derived ages of $60\pm20$~Myr for two clusters M82-F and L located 440~pc south-west of the nucleus; photometric age-dating of the extended region B, located 0.5--1~kpc north-east of the nucleus, shows the peak epoch of cluster formation occurred $\sim$150~Myr ago \citep{smith07}; and spectroscopic observations of 49 clusters throughout the disk show a peak in cluster formation $\sim$140~Myr ago \citep{konstantopoulos09}.

M82 hosts a $\sim$1~kpc long stellar bar, known from near-infrared and H\one\ studies \citep{larkin94, achtermann95, wills00}. This may also have formed as a consequence of the M81-M82 interaction \citep[e.g.][]{barnes96}.
Subsequent to the aforementioned disk-wide burst, it is the action of the bar that has presumably helped to funnel gas into the nuclear regions to fuel the latest starburst episode. Using evolutionary synthesis models, evidence has been found of two recent bursts in the central $\sim$500~pc occurring at $\approx$10 and 5~Myr ago \citep{rieke93, forster03}, and the presence of a strong IR continuum and large CO absorption index indicates a population of red supergiants (RSG) \citep{rieke93, satyapal97, forster01}, further supporting this star formation history.

Of all the starburst clumps, region A is of special interest. It contains a remarkable complex of super star clusters (SSCs) with very high continuum and emission line surface brightnesses \citep{oconnell78, oconnell95}. However, the central regions of M82 notoriously suffer from crowding issues and strong, highly variable obscuration owing to our inclined viewing angle \citep[$i \sim 80^{\circ}$;][]{lynds63, mckeith95}. The high spatial resolution of \textit{HST} is of great benefit for identifying and further studying individual clusters through this very patchy foreground screen.

In \citet[][hereafter \citetalias{smith06}]{smith06}, we presented \textit{HST}/Space Telescope Imaging Spectrograph (STIS) spectroscopy of five massive star clusters in the M82 starburst of varying ages. The main focus of this study was a bright, isolated, SSC in region A designated M82-A1. We determined the age and reddening of M82-A1 using synthetic spectra from population synthesis models and found an age of $6.4\pm0.5$~Myr, meaning it is a product of the most recent starburst event \citep[4--6~Myr ago;][]{forster03}. From \textit{HST} imaging, we also derived a photometric mass estimate of $M =$~7--$13\times10^5$~\Msol, and found it is elliptical with an effective radius of $3.0\pm0.5$~pc and surrounded by a compact ($r = 4.5\pm0.5$~pc) H\two\ region at high pressure ($P/k=1$--$2\times10^7$~cm$^{-3}$~K). The equally high pressures in the surrounding ISM found by \citet[][hereafter \citetalias{westm07c}]{westm07c} led us to conclude that these conditions may have caused the cluster wind to stall \citep[or stagnate; see][]{silich07}.

Here we present a study of the two neighbouring clusters to A1, which we designate A2 and A3\footnote{Clusters 3N, 14N and 61N from \citet{mayya08} and 1a, 1c and 1b from \citet{mccrady07}.}, using \textit{HST}/STIS optical spectroscopy and imaging and Gemini/NIFS near-IR IFU spectroscopy. We find that the age of the three clusters straddles the important epoch of the onset of the RSG phase at $\sim$6~Myr. This sharp turn-on of RSG features represents an extremely accurate age-dating tool, which, if properly calibrated, will be very important in the era of high sensitivity and/or wide field near-IR instrumentation e.g.\ VLT/KMOS or \textit{JWST}.

\section{Observations and data reduction} \label{obs}

\subsection{\textit{HST}/STIS spectroscopy} \label{sect:stis_obs}

\begin{figure*}
\includegraphics[width=\textwidth]{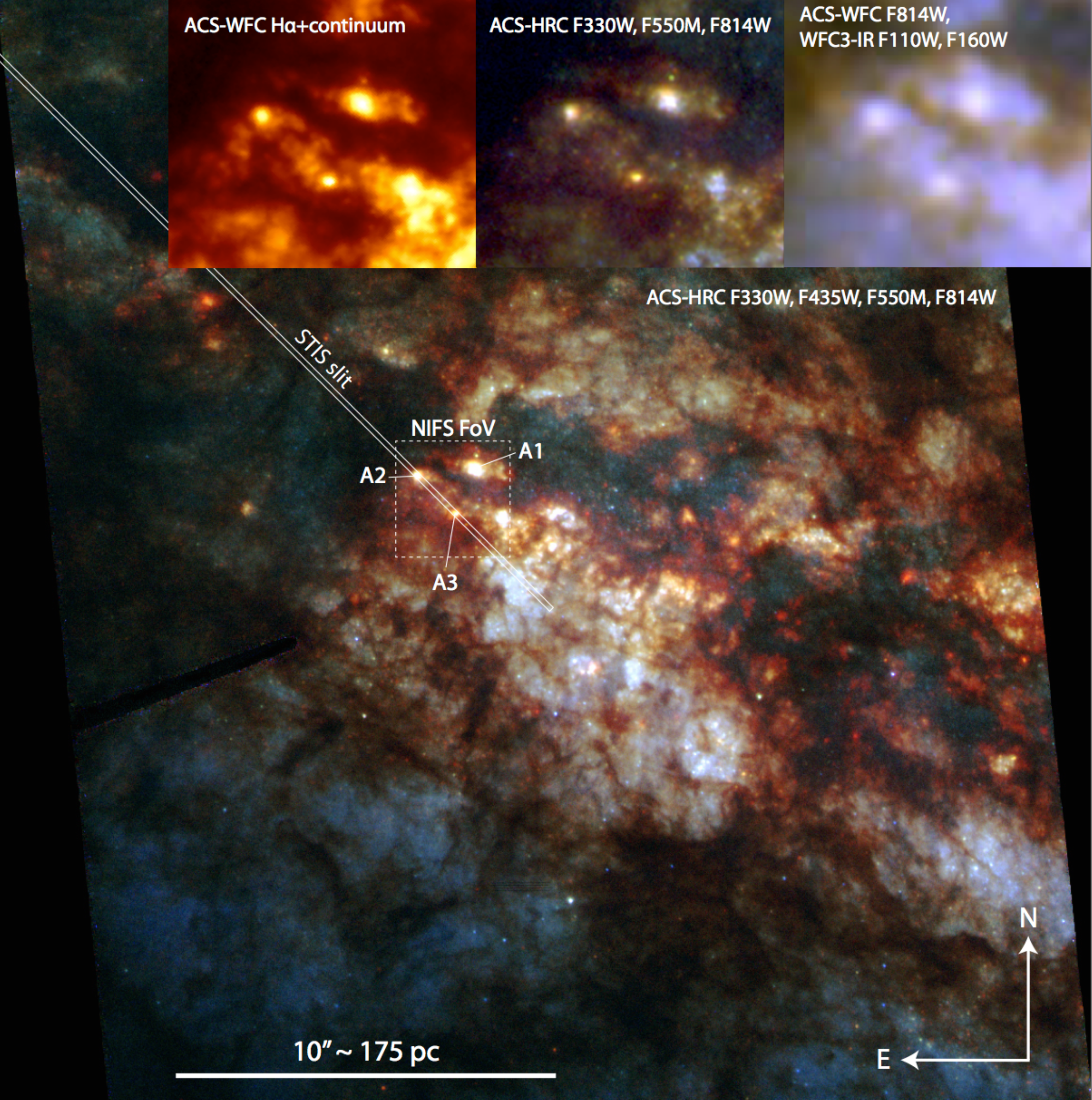}
\caption{\textit{HST}/ACS-HRC colour composite (F330W, F435W, F550M, F814W) of region A of the M82 starburst. The clusters A1 \citepalias{smith06}, A2 and A3 are labelled, together with the location of the $52''$$\times$$0\farcs1$ STIS slit and the field-of-view (FoV) of the NIFS IFU position. The insets at the top show three cut-outs of the A1,2,3 area with the filters as labelled (ACS-WFC observations from program 10776, ACS-HRC from program 10609, and WFC3-IR from program 11360).}
\label{fig:finder}
\end{figure*}

We obtained long-slit \textit{HST}/STIS spectra of clusters M82-A2 and A3 (GO 11641; P.I.\ M.\ Westmoquette) using the G430L and G750M gratings. We utilised the 52$\times$0.1E\footnote{The 52$\times$0.1E aperture centers the target in row 900 of the CCD ($\sim$5$''$ from the end of the slit) close to the readout amplifier in order to mitigate the effects of CTE losses.} aperture at a position angle of 45$^{\circ}$ to cover both the clusters simultaneously. The proximity of the much brighter cluster A1 \citepalias{smith06} meant that we had to initially center the slit (peak-up) on this object, then perform a blind offset to the coordinates of A2 ($\alpha =09^{\rm h}\,55^{\rm m}\,53^{\rm s}.61$; $\delta = +69^\circ\,40'\,50\farcs10$, J2000). We used a 4-point non-integer pixel ($0\farcs415$) dither pattern along the length of the slit to improve the accuracy of spectrum extraction and elimination of hot pixels in the data reduction process. In Fig.~\ref{fig:finder}, we show the slit position on an \textit{HST} Advanced Camera for Surveys (ACS) High Resolution Channel (HRC) colour composite of region A. Details of the observations are listed in Table~\ref{tbl:obs}.

\begin{table*}
\caption{\textit{HST}/STIS Spectroscopic Observations}
\label{tbl:obs}
\centering
\begin{tabular}{ccccc}
Grating & $\lambda$ range (\AA) & $\Delta\lambda$ (\AA\ pix$^{-1}$) & No. of exposures & Total exp time (s) \\
\hline
G430L & 2880--5630 & 2.73 & 42 & 9345 \\
G750M & 6480--7050 & 0.56 & 14 & 3045 \\
\end{tabular}
\end{table*}

The observations were reduced using the \textsc{calstis} pipeline within the \textsc{stsdas} \textsc{iraf} package, after removing from the raw data the known herring-bone noise pattern \citep{jansen03}. This produced wavelength and flux calibrated 2-dimensional spectra for both the G430L and G750M gratings. Next, each of the dithered frames was registered, shifted and then combined using \textsc{ocrrej}, with the associated removal of hot pixels and cosmic-rays. Shifts were determined by measuring the peak of the cluster continuum profile using the \textsc{imexam} `k' command. The resulting reduced data cover the wavelength ranges 2880--5630~\AA\ (G430L) and 6480--7050~\AA\ (G750M).
\begin{figure*}
\includegraphics{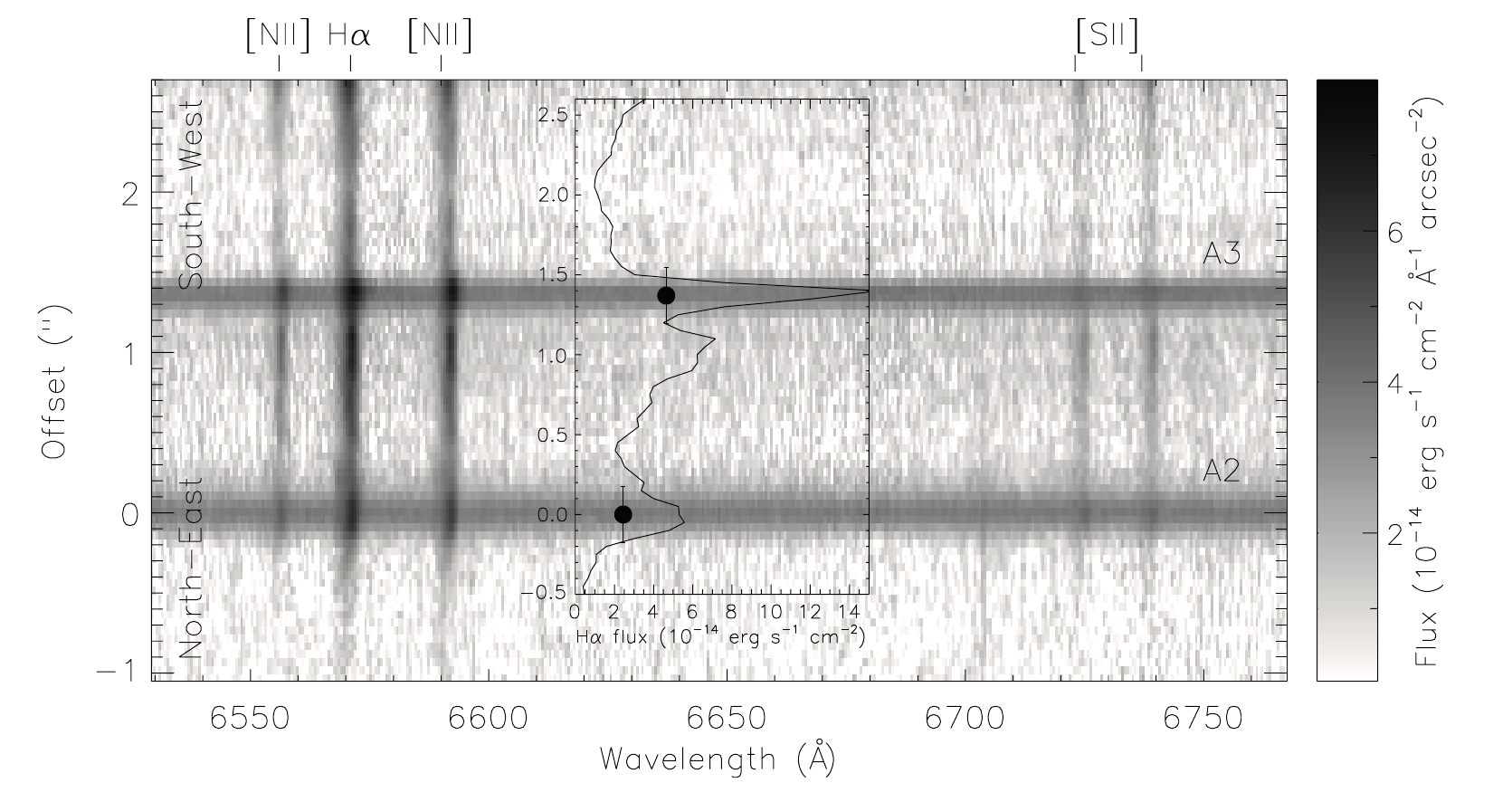}
\caption{STIS G750M two-dimensional spectral image showing the cluster M82-A2 (offset = $0''$) and A3 (offset $\approx$ $1\farcs4$) over the wavelength range 6500--6800\AA. The y-scale is in arcsecs with increasing numbers towards the south-west. The nebular emission lines are marked along the top. The faint vertical feature at $\sim$6703~\AA\ is a bad column. The inset plot shows the integrated H$\alpha$ flux measured every pixel ($0\farcs05$) along this section of the slit and the underlying emission flux variation (i.e.\ without the contribution from the continuum).  Also plotted are the continuum flux measurements near H$\alpha$ from the individual cluster spectra (solid points) with the  7 pixel extraction boxes indicated.}
\label{fig:2dimage}
\end{figure*}

To extract the final one-dimensional spectra of the two clusters, we used the \textsc{calstis} routine \textsc{x1d} using the default extraction box width of 7 pixels for both clusters. Background subtraction was achieved using a 30-pixel-wide region to the north-east of cluster A2, where the nebular emission is negligible compared to the cluster. Using \textsc{x1d} ensures that wavelength dependent aperture illumination effects are taken into account during flux calibration of the spectrum, and does not resample the data in the wavelength axis. 

In Fig.~\ref{fig:2dimage}, a small portion of the reduced two-dimensional image for the G750M grating is shown. Nebular emission is present along the length of this region of the slit and there are distinct peaks at the positions of the two clusters, suggesting that they have H\two\ regions. Indeed, the emission line velocities are slightly offset from the ionized gas seen between the two clusters. These measurements are presented in Sect.~\ref{sect:rvs}.
To measure the spatial profile and extent of the nebular emission associated with the two clusters we measured the H$\alpha$ emission flux from spectra extracted at every pixel along the slit with the continuum subtracted. These measurements are overplotted on Fig.~\ref{fig:2dimage}. The  continuum flux measurements for the clusters near H$\alpha$ and the 7-pixel width extraction box are shown for comparison.
Simple Gaussian fits to the H$\alpha$ spatial profiles of the two clusters gives a FWHM of 5.4~pixels (= 270~mas $\approx$ 4.7~pc) and 2.3~pixels (= 110~mas $\approx$ 2.0~pc) for A2 and A3, respectively, demonstrating that both clusters are embedded in compact H\two\ regions.


The spectral resolution of the G430L and G750M gratings for an extended source is 2--3 pixels. We measure a resolution of 2.42~pixels from Gaussian fits to the nebular emission lines in a G430L spectrum  extracted from a region away from the star clusters (and thus unaffected by Balmer absorption). We therefore adopt a spectral resolution of 2.4~pixels, or 6.6~\AA\ (G430L) and 1.4~\AA\ (G750M grating), in excellent agreement with what we found in \citetalias{smith06}.

\subsection{\textit{HST} imaging} \label{sect:imaging}

\textit{HST} broad and narrow band images were obtained from the Hubble Legacy Archive\footnote{http://hla.stsci.edu/hlaview.html}.  To carry out photometry on the clusters (Section~\ref{sect:photom}) and size measurements
 (Section~\ref{sect:sizes}), we used images obtained with the High Resolution Channel (HRC) of the Advanced Camera for Surveys (ACS) with the F330W (U), F435W (B), F550M (V), and F814W (I) filters (P.I.\ Vacca, PID 10609). These HRC images have a plate scale of $0\farcs027$ per pixel.

\subsection{Gemini/NIFS spectroscopy} \label{sect:nifs_obs}

We obtained Gemini-North Near-Infrared Integral Field Spectrograph (NIFS) observations of the M82-A1, A2 and A3 region on 21st April 2011 (program GN-2010B-Q-4, P.I.\ N.\ Bastian) in seeing-limited mode (AO was not possible due to elevation limits and no natural guide sources were available within or near M82). We used the H-band grating to provide spectra over the wavelength range of 1.48--1.80~\micron\ at a resolution of R=5290 over a 3$''$$\times$3$''$ field-of-view (FoV). The NIFS FoV is shown in Fig.~\ref{fig:finder}. We observed in an ``ABAABA'' object-sky sequence with a 4-point integer-spaxel on-source dither pattern, giving a total on-source exposure time of 2520~seconds. Contemporaneous calibration frames and telluric standard star observations (of HIP52877) were also obtained.

The data were reduced using the Gemini v.1.9 \textsc{iraf} package, utilising pipelines based on the \textsc{nifsexamples} scripts. To obtain a clean telluric absorption spectrum from the HIP52877 observations, the spectra were processed as described for the science images below, after which 1-D spectra were extracted and the Brackett (Br) absorption lines were fitted and removed.

Based on the methodology detailed in \citet{seth10}, each science exposure was sky subtracted using the nearest off-source exposure, then sliced up and rectified based on Ronchi mask and arc lamp images using a custom version of \textsc{nftransform} that propagates the VAR and DQ extensions. Each frame was then corrected for telluric absorption using a custom version of \textsc{nftelluric}, and spatially rebinned into data-cubes using a custom IDL script based on the \textsc{nifcube} routine, again to preserve the VAR and DQ extensions. In this process, the original 0\farcs043$\times$$0\farcs1$ spaxels were rebinned into $0\farcs05$ spaxels. The final data-cubes from each individual on-source exposure were combined using a custom IDL script, correcting for the spatial offsets. Based on measurements of the sky lines, the spectral resolution is $3.19\pm0.35$~\AA\ (60~\kms). The spatial resolution of the combined image was measured by fitting a 2-D Gaussian to the collapsed telluric star data-cube and found to be FWHM=$0\farcs49$.

\section{Measuring the cluster properties}

We utilize the observations and data described above to derive a number of properties of the clusters under study. These include age, extinction, radial velocities, size, and mass, as described in the following sections.
\begin{figure*}
\centering
\includegraphics[width=0.7\textwidth]{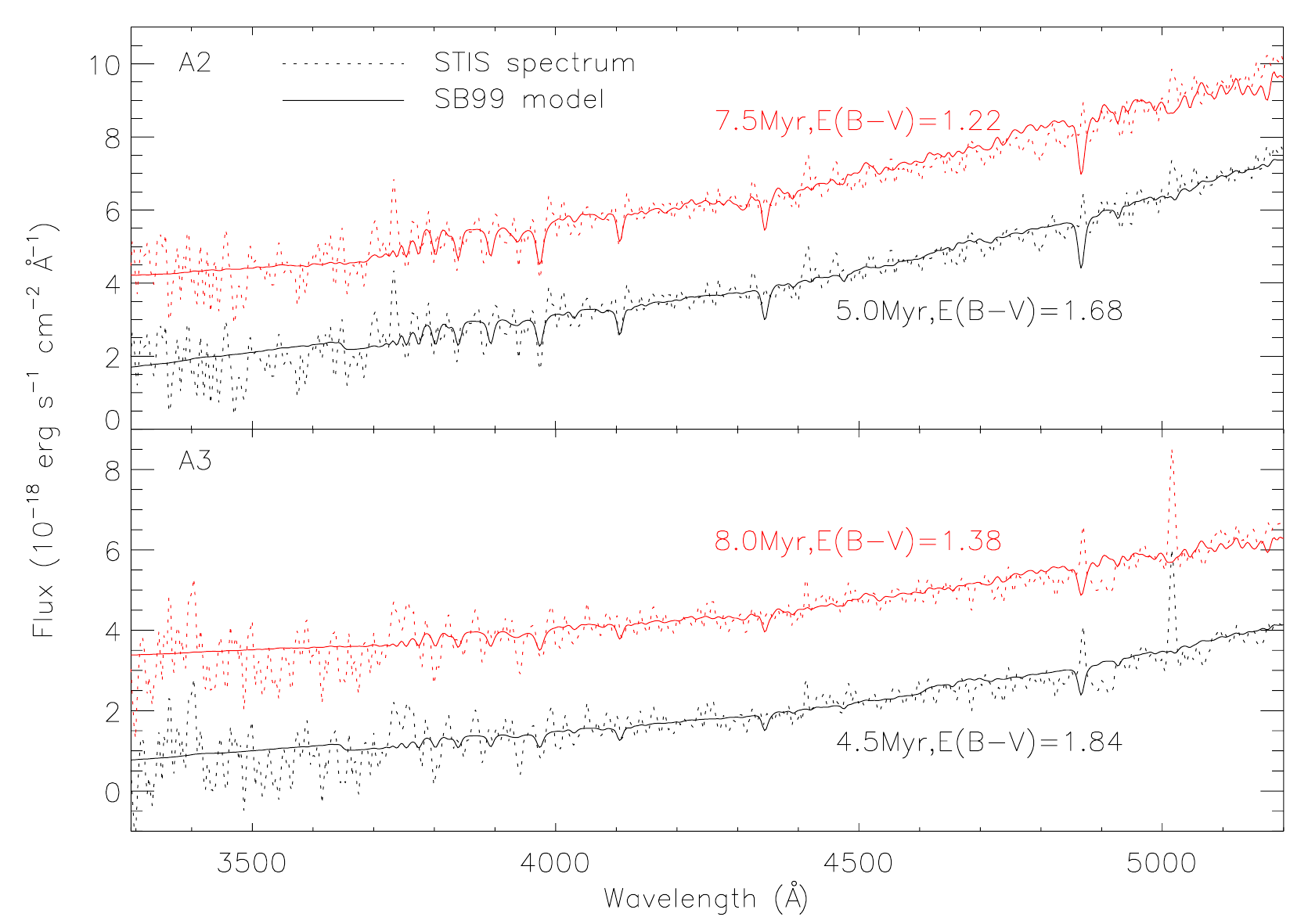}
\caption{Comparison of the reddened SB99 model spectra for two of the best fitting ages and extinctions to the STIS observations of A2 and A3. Nebular emission line features in the observed spectra were masked out in the fitting process.}
\label{fig:stis_spectra}
\end{figure*}
\subsection{Extinctions and ages from STIS spectra}\label{sect:sb99_fits}
We adopt the methods used in \citetalias{smith06} to determine the extinction and age of clusters A2 and A3. This method compares synthetic spectra generated with the evolutionary synthesis code Starburst99 \citep[SB99;][]{leitherer99} to the observed spectra using the continuum slope longwards of the Balmer jump to derive the reddening and the region below the Balmer jump to provide the age.

We binned the G750M STIS spectrum to match the G430L spectrum and merged the two spectra. We then generated a series of SB99 models (version 6.03) over the age range of 1--12~Myr in steps of 0.5~Myr, assuming solar metallicity,
an instantaneous burst with a \citet{kroupa01} IMF, lower and upper masses of 0.1 and 100~\Msol, and the enhanced mass-loss Geneva tracks. The resulting high resolution synthetic spectra were smoothed, binned, and velocity shifted to match the G430L spectra. In \citetalias{smith06}, we showed that the strengths and ratios of the nebular [N\two] and [O\two, \three] lines in the H\two\ region associated with cluster A1 are consistent with solar metallicity.

We reddened the resulting synthetic spectra using a foreground dust screen, employing the Galactic reddening law from \citet{howarth83} with $R=A(V)/E(B-V)=3.1$, and a range of $E(B-V)$ values from 1.0--2.2 mag in steps of 0.02 mag. These reddened spectra were then normalised to the observed cluster spectra over 6850--6950~\AA\ and compared. We chose to normalise at the reddest wavelengths to provide maximum leverage for determining the reddening from the spectral slope. The quality of the fit for each value of $E(B-V)$ was judged by the $\chi^2$ statistic over the relatively featureless wavelength range 4140--5200~\AA.  Any strong emission or absorption features were masked out in the fitting process for the continuum. We chose to redden the synthetic spectra rather than de-redden the observed spectra to keep the noise in the observed spectra constant. Using this approach, and examining the fits visually, we find that the reddening can be tightly constrained to $\pm 0.05$~mag for a given age.

To obtain age estimates, we compared the wavelength region between 3300--4300~\AA\ in the observed and reddened synthetic spectra for those reddening values obtained from the continuum fits for each age in the range 1--12~Myr. We preferred to split the $\chi^2$ fitting into two parts because the noise below the Balmer jump dominates the quality of the fit. Indeed, this level of noise limits the accuracy of the age determination. The data are far too noisy below 3300~\AA\ for any useful analysis.

From the $\chi^2$  fits for M82-A2 and M82-A3, we find that ages of 3~Myr or younger give poor fits because the continuum blueward of the Balmer jump is too strong. Above this age, we find that a unique age cannot be assigned using the goodness-of-fit criterion because the data are too noisy near the Balmer jump. In Fig.~\ref{fig:stis_spectra} we show fits for M82-A2 for 5.0 Myr, $E(B-V)=1.68$ and 7.5~Myr, $E(B-V)=1.22$, and for M82-A3 for 4.5~Myr, $E(B-V)=1.84$ and 8.0~Myr, $E(B-V)=1.38$. Clearly, the two ages and reddenings plotted give equally acceptable fits for the respective clusters. We note that we employed the same methods described above for the M82-A1 spectra and recovered the narrow range in age of 6--7~Myr and a reddening of $E(B-V)=1.3$~mag \citepalias{smith06}. The longer exposure time for the M82-A1 observations coupled with the lower reddening give a much cleaner spectrum below the Balmer jump and this acts as an effective age discriminant. 

To put constraints on the upper age limit for M82-A2 and A3, we consider the nebular diagnostic lines in the STIS spectra. We note that the presence of ionized gas (Fig.~\ref{fig:2dimage}) suggests ages of $<$10~Myr, and the clear detection of [O\three]$\lambda5007$ in the spectra (Fig.~\ref{fig:stis_spectra}) suggests even younger ages. The STIS data are too noisy to use the Wolf-Rayet emission line feature at 4700\AA\ as an age discriminant (weak WR emission is seen in the model spectrum at 4.5~Myr in Fig.~\ref{fig:stis_spectra}). We conclude from modelling the STIS spectra with evolutionary synthesis models that the probable ages of M82-A2 and A3 are in the range 3.5--10 Myr. In the next section, we show how the age-reddening degeneracy in the STIS spectra can be broken using the near-infrared data.


%
\subsection{Radial Velocities from STIS spectra}\label{sect:rvs}
We explored various methods for obtaining the radial velocities of the clusters. The lack of features in the H-band spectra (next section) and the low signal-to-noise and resolution of the G430M STIS spectra prevented us from using spectral absorption lines for radial velocity measurements. We therefore used the emission lines associated with the compact H\two\  region surrounding each cluster to provide velocity measurements. Gaussian profiles were fitted to the H$\alpha$ and the [N\two] doublet and yielded mean velocities of $330\pm6$ (A2) and $336\pm5$  \kms\ (A3). For comparison, the ionized gas between these two clusters has a velocity of $323\pm2$ \kms. This difference in emission line velocities between the clusters and the ionized gas suggests that the H\two\ regions are indeed associated with the clusters, and can be used as a good proxy for the cluster radial velocities. This is strengthened by our finding in Paper~I for cluster A1 that the mean velocities of the Balmer absorption lines and nebular emission lines are the same: $319\pm22$ \kms\ and $320\pm2$ \kms.

\subsection{H-band cluster spectra}\label{sect:hband}

\begin{figure}
\centering
\includegraphics[width=0.45\textwidth]{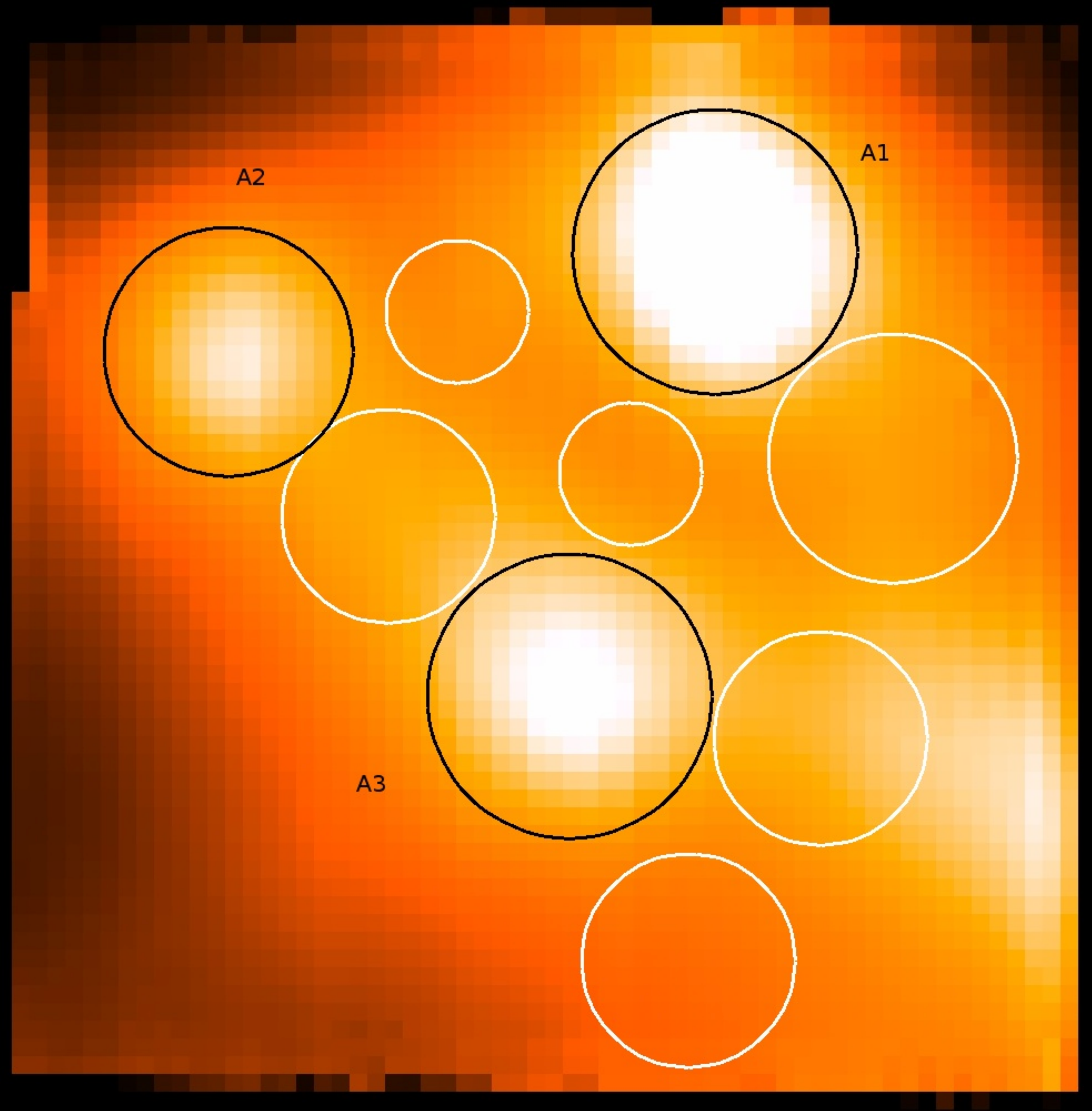}
\caption{NIFS H-band continuum image showing the apertures (black circles) and background regions (white circles) used to extract spectra from the data-cube.}
\label{fig:nifs_apertures}
\end{figure}

\begin{figure*}
\centering
\includegraphics[width=\textwidth]{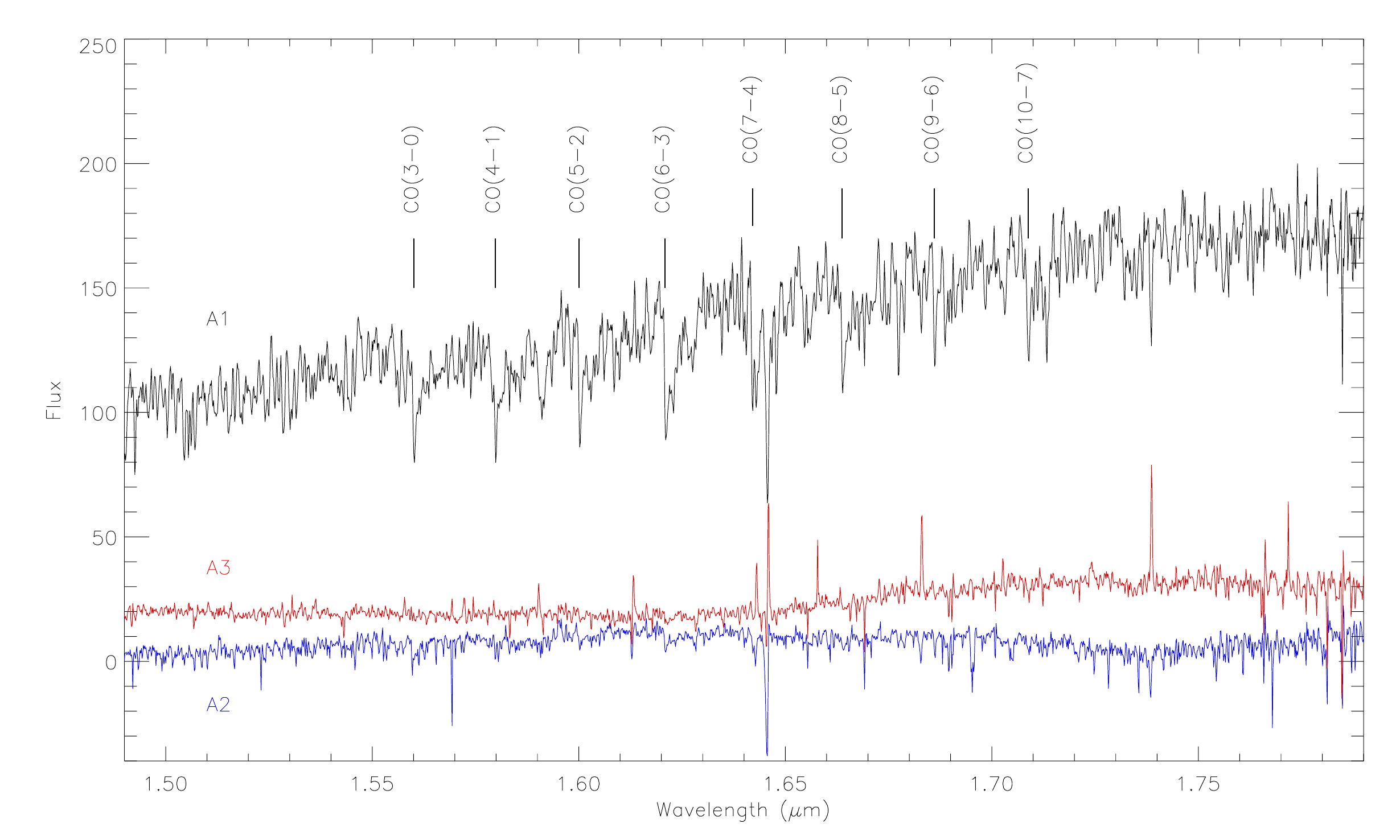}
\caption{H-band background-subtracted spectra of the three clusters using the apertures shown in Fig.~\ref{fig:nifs_apertures} (arbitrary but relative flux units). The $^{12}$CO bandheads clearly seen in the A1 spectrum are labelled in black \citep[identifications from][]{rayner09}. 
}
\label{fig:nifs_spectra}
\end{figure*}

We extracted H-band spectra of clusters A1, A2 and A3 from the NIFS datacube using object apertures with radii of 0\farcs40 (A1, A3) and 0\farcs35 (A2).
To subtract the background, we experimented with various apertures and chose to use the average of six apertures with sizes of 0\farcs20--0\farcs35, as shown in Fig.~\ref{fig:nifs_apertures}.
We used offset background apertures rather than annuli, since these gave cleaner resulting spectra due to the variable brightness of the background. The background-subtracted cluster spectra are shown in Fig.~\ref{fig:nifs_spectra}. 



What is immediately obvious from the spectra is the presence (or absence) of the $^{12}$CO absorption bandheads. A1 exhibits strong CO absorption, whereas in both A2 and A3 these absorption lines are weak or absent. It should be noted that the background spectrum contains CO absorption features from the underlying stellar population. The very weak or absent CO absorption in A2 and A3 was recovered for any combination of background aperture subtractions. We thus suspect that any residual CO absorption in these two clusters most likely arises from the background, which contains red supergiants \citep{greco12}.
Contributors to CO bandheads in integrated cluster spectra are pre-main-sequence (PMS) stars \citep{meyer05} and red supergiants (RSGs). Above an age of $\sim$1~Myr, the PMS contribution diminishes. Above this, where RSG stars exist, they dominate the light output in the near-IR. For example, a single RSG would contribute $>$90 percent of the total light emitted in the J-band for a cluster of mass $10^4$~\Msol\ \citep{gazak13}. Thus for the masses of A2 or A3, $\sim$3 or 4 RSGs are needed to dominate the near-IR luminosity. The expected age for the appearance of the first RSG (based on a wide variety of stellar evolutionary models) is $5.7\pm 0.8$~Myr \citep{gazak13}. As shown by \citet{figer06} the near-IR CO bandhead absorption features become stronger for later spectral types, and are particularly strong for supergiants. This means they are a powerful diagnostic for the presence of RSGs.

From our SB99 model fitting of the STIS spectra, we find a degeneracy in the age solution for clusters A2 and A3 where a range from 3.5--10~Myr is equally acceptable, whereas A1 has a clear solution at 6--7~Myr. However, Fig.~\ref{fig:nifs_spectra} shows that A1 exhibits strong CO absorption bands and therefore must contain RSG stars, whereas  in A2 and A3, they are weak or absent. We therefore conclude that clusters A2 and A3 are probably in the younger (pre-RSG phase) age range of $4.5\pm1.0$~Myr, and that these two, together with the slightly older cluster A1, straddle this important step in cluster evolution. We have caught A1 at the point where the first RSGs have appeared. This technique represents an extremely accurate age-dating tool, and  verifies the ages of these three clusters.

\citet{mccrady07} presented Keck/NIRSPEC spectra at $R=22$\,000 for clusters 1a ($\equiv$A1) and 1c ($\equiv$A2). With these they derived their dynamical masses; they found A1 to have a mass of 8.6($\pm$1.0)$\times10^5$~\Msol\ and A2 to be 5.2($\pm0.8$)$\times 10^5$~\Msol. Their spectra show CO absorption features for both clusters (whereas we only find them for A1), however their ground-based, long-slit observations are likely to have suffered from some level of contamination due to their poorer seeing conditions ($0\farcs7$ vs.\ $0\farcs5$ for our NIFS observations) and from a more limited quality background subtraction since they did not have the advantage of IFU coverage to select the most appropriate background.

\subsection{Photometry} \label{sect:photom}
We performed aperture photometry on clusters A1, A2, A3 using the ACS/HRC images as described in Section~\ref{sect:imaging}. We used a fixed aperture size of $0\farcs35$ for all the filters, along with a background annulus and inner radius of $0\farcs5$ with a width of $0\farcs1$. Counts were converted to magnitudes using the photometric \textsc{vegamag} zeropoints given on the relevant STScI ACS webpages; aperture corrections of 0.3~mag were applied to each cluster in all filters. We estimate that the errors on the photometry amount to $\sim$0.1~mag, where the largest uncertainty is due to the highly variable background around the clusters. Our results are given in Table~\ref{tbl:obs-images}.

\begin{figure*}
\centering
\includegraphics[width=0.7\textwidth]{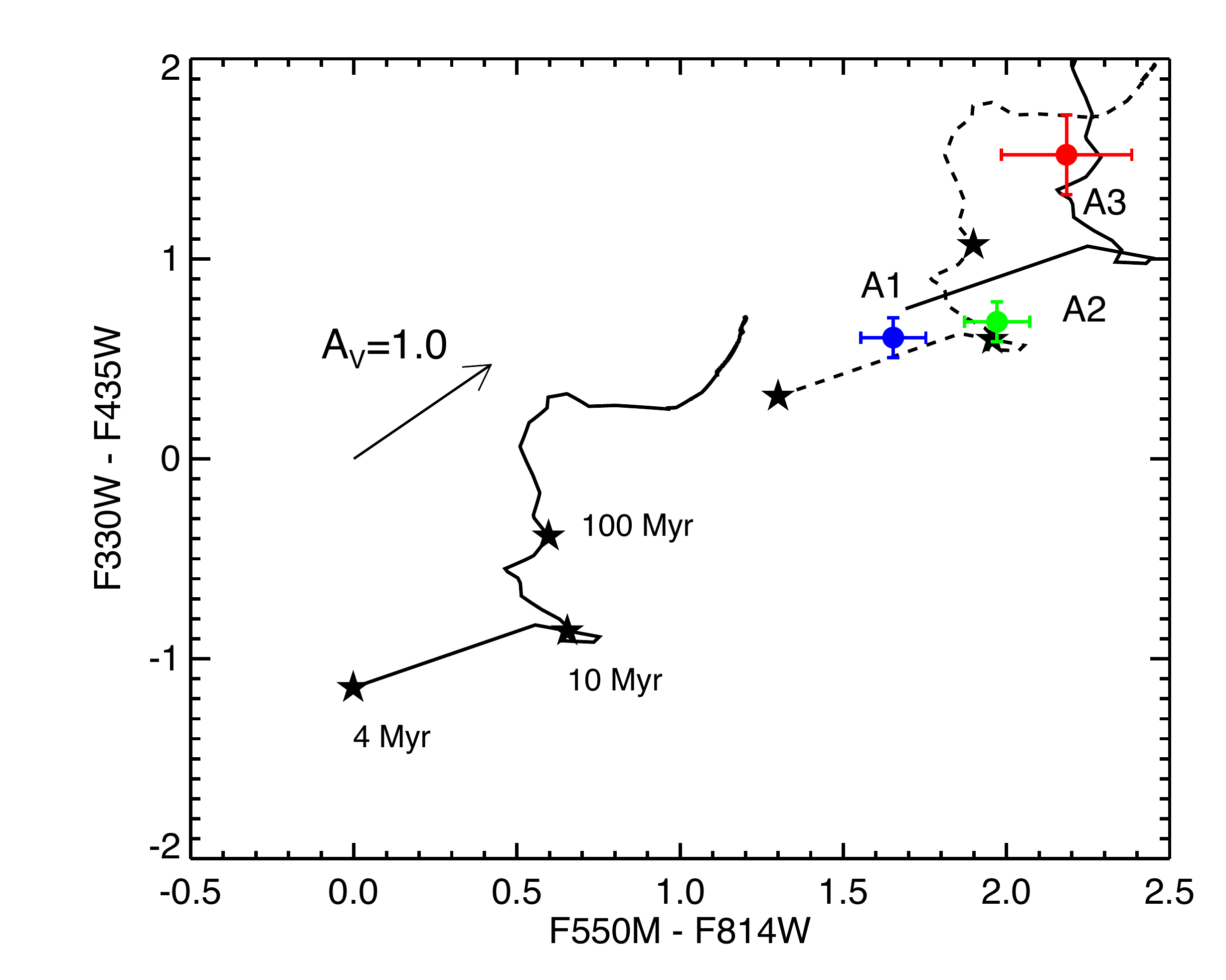}
\caption{Color-color plot for clusters A1, A2 and A3.
The solid line represents the solar metallicity GALEV model \citep{kotulla09} from 4~Myr to 16~Gyr, where the filled stars show ages of 4, 10 and 100 Myr. The dashed line shows the same model shifted by $E(B-V) = 1.0$. }
\label{fig:cmd}
\end{figure*}

In Fig.~\ref{fig:cmd} we plot the cluster A1, A2 and A3 photometry in color-color space. We overplot the GALEV evolutionary synthesis models \citep{kotulla09} from 4~Myr to 16~Gyr for solar metallicity.  M82-A1 is consistent with an age of 6.5~Myr and $E(B-V) \approx 1.0\pm0.3$,
in good agreement with \citetalias{smith06}. The best-fitting photometric extinctions for clusters A2 and A3 are $E(B-V) \approx 1.3\pm0.3$ and
$E(B-V) \approx 1.5\pm0.5$.
Photometric ages (see Fig.~\ref{fig:cmd}) are not in good agreement with those derived from the optical spectral fitting described above. This is likely due to the high amount of extinction present (causing large corrections to the photometry), a  as well as the presence of differential extinction across the face of each cluster \citep[e.g.][]{bastian07}. Additionally, we note that the (deconvolved) size of all three clusters increases to redder wavelengths (see Section~\ref{sect:sizes} below), suggesting lost flux in bluer filters, causing artificially redder colours. This is due to the fact that the outer regions of the cluster fall below the background level when there is high extinction \citep[or highly variable extinction; e.g.][]{gall_smith99}. As such, we will only use the photometry to estimate the mass of each of the clusters.

Using the spectroscopically derived ages and extinctions with the F550M photometry, we determine the absolute V-band magnitudes, $M_{V}$, and photometric masses using the GALEV models and a Kroupa IMF \citep{kroupa01}. These values are given in Table~\ref{tbl:cluster_params}. The resulting masses of the three clusters are surprisingly similar, in the range 4--$7\times10^5$~\Msol. Errors of order 2--$3\times10^5$~\Msol\ take into account the uncertainties on all the input factors, including the form of the IMF. These results are in excellent agreement with those derived in \citetalias{smith06}, by \citet{mayya08}, and the dynamical masses derived by \citet{mccrady07} as mentioned above.

\subsection{Sizes} \label{sect:sizes}
To measure the size of M82-A2 and A3, we ran the {\sc ishape} algorithm \citep{larsen99b} on the clusters in each of the four individual FLT frames obtained for each filter of the ACS-HRC imaging. We decided not to use the combined (and hence cosmic-ray cleaned) drizzled images since it is very difficult to obtain a reliable and accurate point-spread function (PSF) for drizzled images where there are not enough stars in the field to create an empirical PSF. Thus we used the {\sc tinytim} package \citep{krist04} to create a model PSF for the ACS-HRC images. Following \citetalias{smith06} we used a Moffat function with a power index of 1.5.

In Table~\ref{tbl:obs-images}, we list the average major axis FWHM and minor/major axis ratios from the fits to the four individual frames. We also give the effective or half-light radius $R_{\rm{eff}}$ where we have used the expressions given by \citet{larsen04} to convert the FWHM of an elliptical Moffat profile fit to a half-light radius. In the V-band, we find A2 to have a size of $R_{\rm eff}=104$~mas ($\sim$1.8~pc) and A3 to have a size of $R_{\rm eff}=59$~mas ($\sim$1.0~pc). For comparison, the ACS-HRC camera pixel scale is 27~mas as detailed in the ACS instrument handbook \citep{ubeda12}. Using these newer HRC images, we also find A1 to have a size of 159~mas ($\sim$2.8~pc), which is in excellent agreement with what we found in \citetalias{smith06} using WFPC2 imaging.

\definecolor{Gray}{gray}{0.9}
\begin{table*}
\caption {\textit{HST} photometry and derived parameters for cluster M82-A1, 2 and 3. The errors in the effective radius and magnitude measurements are $\pm30$~mas and $\pm0.10$ mag respectively (except for A3 where the photometic error is $\pm0.2$ mag).}
\label{tbl:obs-images}
\centering
\begin{tabular}{@{}llcccccc}
Filter & Detector & Exposure time & FWHM & Minor/Major & $R_{\rm{eff}}$& {\sc vegamag} \\
&  & (s) & (major axis) & Axis Ratio &  (mas) & Photometry & \\
&& & (mas) && & (mag)\\
\hline 
\textbf{A1}\\

F330W & ACS-HRC  & 4736 & 130.8 & 0.93 & 143 & 19.47\\
F435W & ACS-HRC  & 1132 & 150 & 0.80 & 153 &  18.87\\
F550M & ACS-HRC  & 840 & 158 & 0.78 & 159 & 17.27\\
F814W & ACS-HRC  & 140 & 241 & 0.77 & 241 & 15.62\\

\hline 
\textbf{A2}\\

F330W & ACS-HRC  & 4736 & 63 & 0.70 & 59 & 20.90\\
F435W & ACS-HRC  & 1132 & 75 & 0.69 & 72 &  20.21\\
F550M & ACS-HRC  & 840 & 108 & 0.71 & 104 & 18.24\\
F814W & ACS-HRC & 140 & 150.0 & 0.86 & 158 & 16.27\\

\hline 
\textbf{A3}\\

F330W & ACS-HRC  & 4736 & 36 & 0.86 & 42 & 22.74\\
F435W & ACS-HRC  & 1132 & 46 & 0.69 & 51 &  21.22\\
F550W & ACS-HRC  & 840 & 63 & 0.67 & 59 & 18.69\\
F814W & ACS-HRC  & 140 & & failed to fit && 16.51\\

\hline
\end{tabular}
\end{table*}

\subsection{Summary of cluster parameters}\label{sect:params}

Table~\ref{tbl:cluster_params} summarises all the measured and derived properties of the three clusters, M82-A1, A2 and A3, determined from this study and that of \citetalias{smith06}. 
From SB99 fits to the optical STIS spectra of A2 and A3 (Section~\ref{sect:sb99_fits}), we find a degeneracy in the age solution for both clusters where a range from 3.5--10~Myr (and reddenings of $E(B-V) =1.2$--1.9~mag) is equally acceptable, whereas A1 has a clear solution at 6--7~Myr (with $E(B-V)=1.3$~mag). From the strength of CO absorption features arising from RSGs in our Gemini/NIFS H-band spectra of the three clusters (Section~\ref{sect:hband}), we were able to break this degeneracy since A1 shows RSG features, whereas in A2 and A3, they are weak or absent. A2 and A3 must therefore be in the younger $4.5\pm 1.0$~Myr age range. We measured their photometry from \textit{HST}/ACS-HRC imaging thus giving their photometric masses (Section~\ref{sect:photom}), their sizes from \textsc{ishape} fits  (Section~\ref{sect:sizes}), and their radial velocities from the nebular emission lines in the STIS spectra (Section~\ref{sect:rvs}).

\begin{table*}
\centering
\caption{Summary of derived parameters for clusters M82-A1, A2 and A3 from this study (except where noted).}
\label{tbl:cluster_params}
\begin{threeparttable}
\begin{tabular}{@{}lr@{}c@{}lr@{}c@{}lr@{}c@{}l}
& \multicolumn{9}{c}{\textbf{Cluster}}\\
Parameter & \multicolumn{3}{c}{\textbf{A1}} & \multicolumn{3}{c}{\textbf{A2}} & \multicolumn{3}{c}{\textbf{A3}}\\
\hline
Radial velocity, $V_{\rm r}$ & 320 & $\pm$ & 2 km s$^{-1}$\tnote{\textdagger}& 330 & $\pm$ & 6 km s$^{-1}$ & 336 & $\pm$ & 5 km s$^{-1}$ \\
F555W & $17.52$ & $\pm$ & $0.10$ mag & $18.57$ & $\pm$ & $0.10$ mag & $19.03$ & $\pm$ & $0.30$ mag\\
Half-light radius, $R_{\rm{eff}}$ & 2.8 & $\pm$ & 0.3 pc & 1.8 & $\pm$ & 0.3 pc & 1.0 & $\pm$ & 0.3 pc \\
$E(B-V)$ & 1.35 & $\pm$ & 0.15 mag\tnote{\textdagger} & 1.70 & $\pm$ & 0.2 mag & 1.85 & $\pm$ & 0.3 mag \\
$M_{V}$\tnote{\textdaggerdbl} & $-14.44$ & $\pm$ & 0.46~mag & $-14.46$ & $\pm$ & 0.61~mag & $-14.48$ & $\pm$ & 0.95~mag \\
Age & 6.4 & $\pm$ & 0.5\tnote{\textdagger} & 4.5 & $\pm$ & 1~Myr & 4.5 & $\pm$ & 1~Myr \\
Mass, $M$ & 5.6 & $\pm$ & $2.8\times10^{5}$ M$_{\odot}$ & 4.0 & $\pm$ &$2.0\times10^{5}$ M$_{\odot}$ & 7.2 & $\pm$ &$3.6\times10^{5}$ M$_{\odot}$ \\
\noalign{\vskip 1ex} Electron density, $N_{\rm e}$ & $1800$ &$^{+}_{-}$& $^{340}_{280}$ \cmt\tnote{\textdagger} & $800$ &$\pm$&150 \cmt & $1030$ & $\pm$ & 150 \cmt\\
\hline
\end{tabular}
\begin{tablenotes}
\item[\textdagger] From \citetalias{smith06}.
\item[\textdaggerdbl] Corrected for extinction.
\end{tablenotes}
\end{threeparttable}
\end{table*}

\section{Discussion}

\begin{figure*}
\centering
\includegraphics[width=0.48\textwidth]{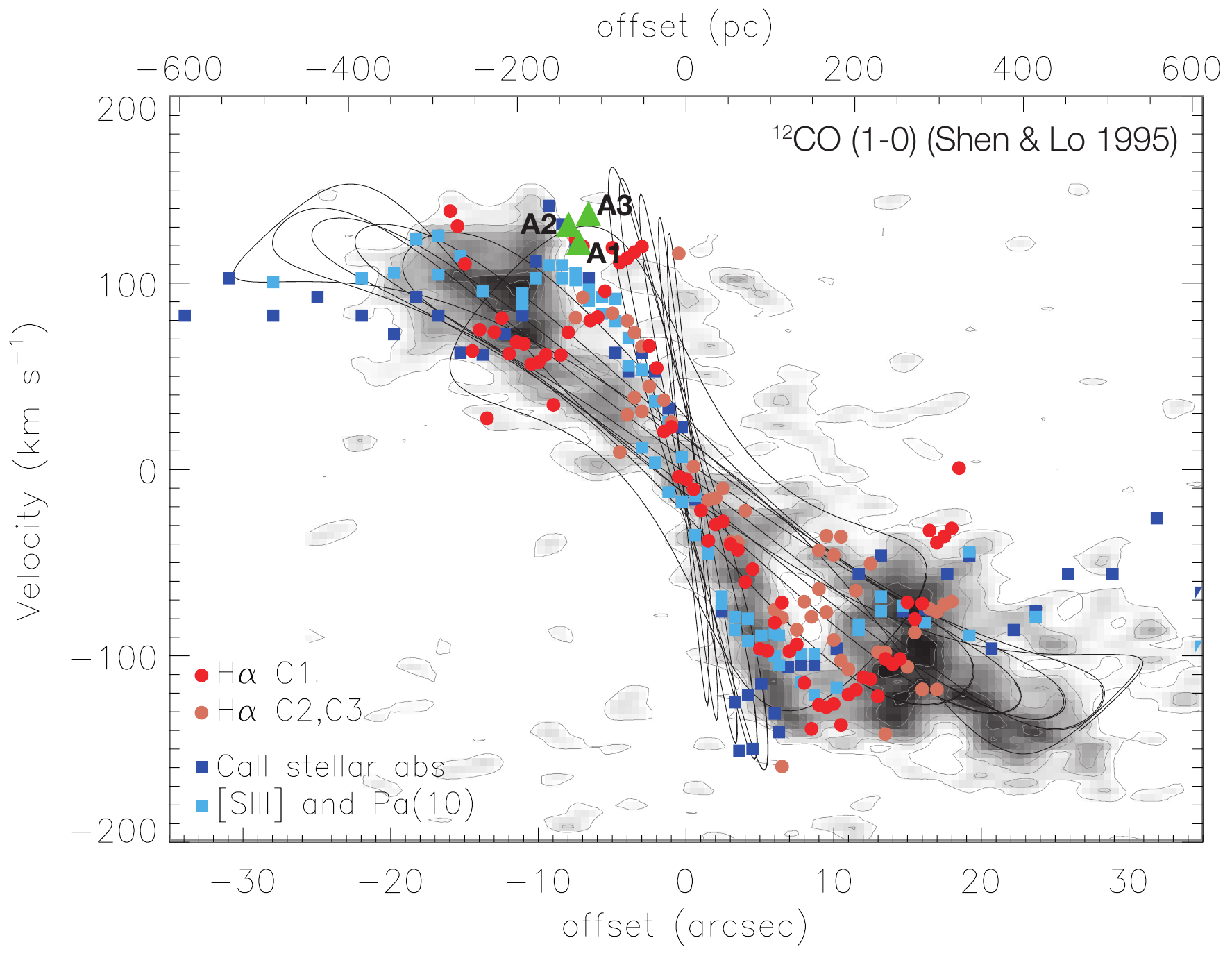}
\includegraphics[width=0.48\textwidth]{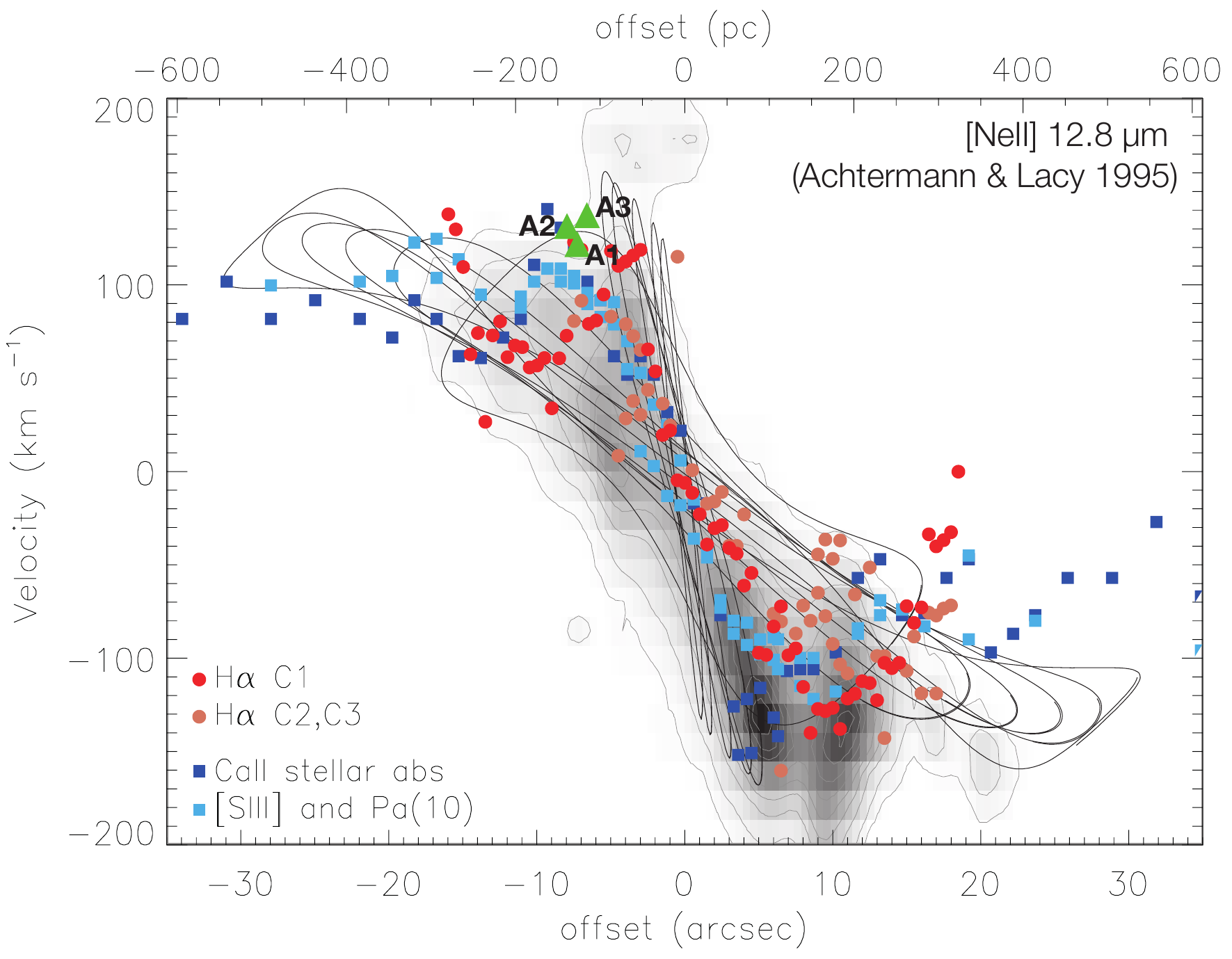}
\caption{Major axis PV diagrams for $^{12}$CO ($J = 1\rightarrow0$) \citep{shen95} and [Ne\two]\,12.8\,\micron\ \citep{achtermann95} (greyscale with contours) with the bar model predictions of \citet{wills00} superimposed (solid line orbit tracks; reproduced from \citealt{wills00}). Also overlaid are the H$\alpha$ \citepalias[C1, C2 and C3 represent different fitted line components;][]{westm07c}, near-IR [S\three]$\lambda$9069 and Pa(10)$\lambda$9014 \citep{mckeith93} and the Ca\two\,$\lambda$8542 stellar absorption line \citep{mckeith93} radial velocities. The green triangles show the emission line radial velocities of the three clusters A1, A2 and A3 (the uncertainties are approximately the size of the plotting symbol; see Section~\ref{sect:rvs}). A physical scale in pc is given at the top.}
\label{fig:x2_models}
\end{figure*}

\subsection{How and where did these clusters form?}

In projection, clusters A1, A2 and A3 form a triplet located $\sim$8$''$ ($\sim$140~pc) to the north-east of the nucleus, in a region of moderate extinction \citepalias[$A_{V}$=4--5.5~mag;][]{westm07c} and high electron densities/pressures \citep[][Section~\ref{sect:environment}]{westm09b}.

\begin{figure*}
\centering
\includegraphics[width=0.8\textwidth]{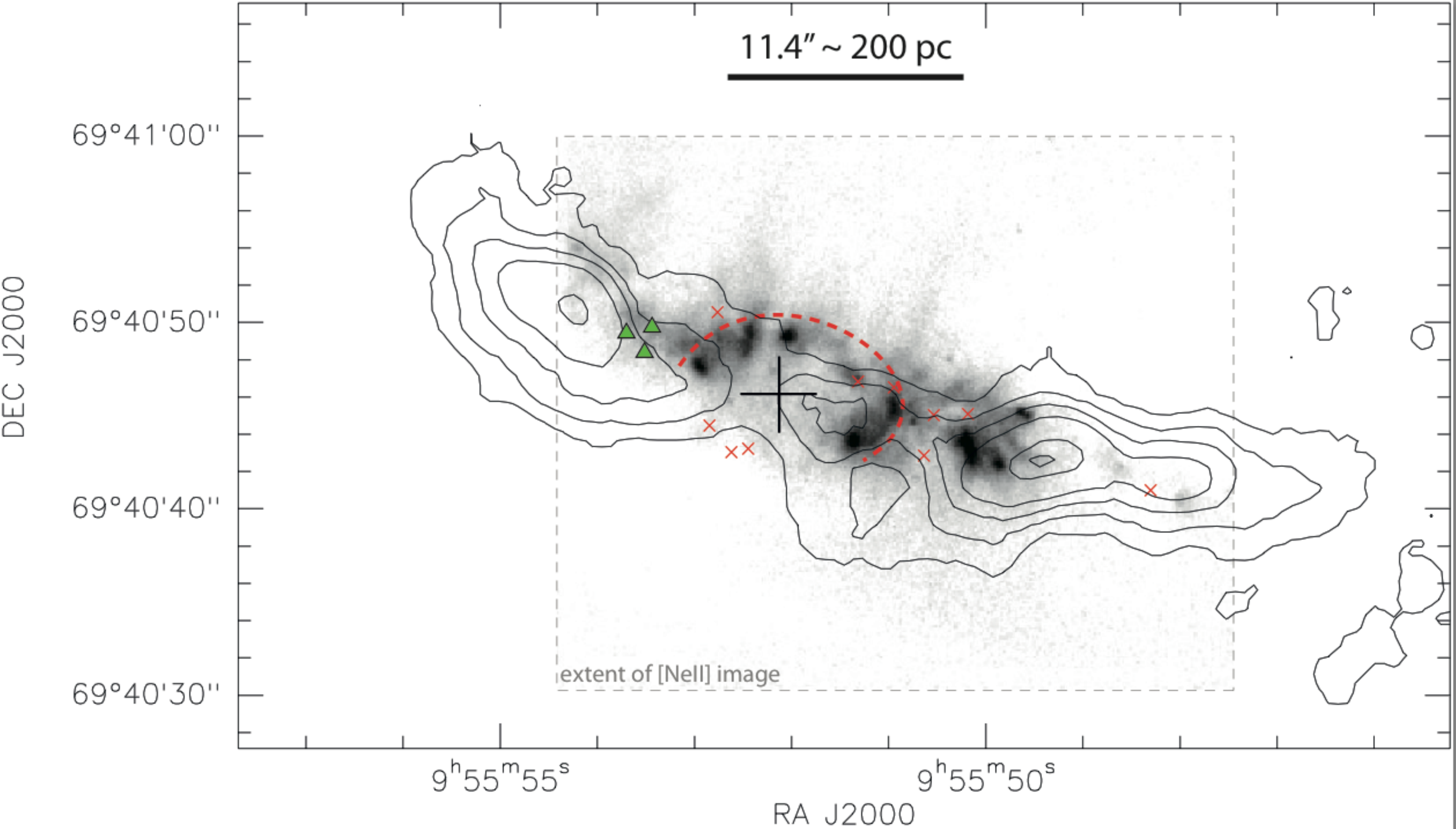}
\caption{The greyscale image (contained within the dashed box) is the Subaru/COMICS [Ne\two]\,12.8~\micron\ emission line image reproduced from \citet{gandhi11}. The overlaid contours represent the integrated C$^{18}$O ($J = 1\rightarrow0$) line intensity from \citet{weis01} observed with the PdBI. The black cross indicates the 2.2~\micron\ nucleus of M82, and the red dashed curve outlines part of the mid-IR circumnuclear ionized gas ring \citep[first proposed by][and clearly seen here]{larkin94, achtermann95}. The location of clusters A1, A2 and A3, indicated with green triangles, fall between the eastern edge of the circumnuclear ring and the eastern side of the molecular torus. The red crosses indicate the SSCs identified by \citet{mccrady03} from near-IR imaging.}
\label{fig:CO+NeII}
\end{figure*}

In Fig.~\ref{fig:x2_models} we present a multi-wavelength comparison of the major axis position-velocity (PV) diagram. We include the optical and near-IR emission line and stellar radial velocities measured in \citetalias{westm07c} and \citet{mckeith93}, together with the $^{12}$CO(2-1) \citep{shen95} and [Ne\two]\,12.8\,\micron\ \citep{achtermann95} measurements, and the bar model predictions of \citet{wills00}. The shallower orbit tracks extending to $\pm$25$''$ represent the $x_{1}$ orbits of the bar, whereas the steeper ones in the central $\pm$5$''$ represent the perpendicular $x_2$ orbits. The emission line-derived radial velocities of the three clusters are shown with green triangles, placing them at the extreme positive velocity end of the major-axis PV diagram.
All three clusters have similar radial velocities, implying that they lie at similar radial distances.
In Fig.~\ref{fig:CO+NeII} we compare the projected location of the clusters with the morphological distribution of the ionized [Ne\two]\,12.8~\micron\ emission \citep{gandhi11} and the integrated C$^{18}$O(1-0) line intensity \citep{weis01}.

These figures highlight two main findings. Firstly the well-known molecular torus \citep[R$\sim$15$''$;][]{shen95, lord96, weis01, fuente08} can be seen as the two bright knots in the CO PV diagram (Fig.~\ref{fig:x2_models}) at $\pm$100~\kms, and also clearly in Fig.~\ref{fig:CO+NeII}. This torus is thought to have formed as a result of the interaction between material on the $x_1$ and $x_2$ orbits: the existence of orbital resonances at certain radii (such as the inner Lindblad resonance; ILR) can prevent gas from flowing past the resonance, causing a ``pile-up''. As the bar pattern rotates, a torus is formed from this material. In these pile-ups, gas can lose angular momentum to the stellar bar as it shocks, and consequently fall radially inwards \citep{jenkins94, buta96}. So-called orbit spraying \citep{athanassoula92a} can also transfer material from the $x_1$ to $x_2$ orbits causing it to shock and form dust lanes and stars \citep{greve02}.
In addition, gravitational torques associated with the bar can also transport gas inwards \citep{vanderlaan11}.

Secondly, the mid-IR circumnuclear ionized gas ring (seen in [Ne\two], Br$\gamma$, and H92$\alpha$; \citealt{larkin94, achtermann95, r-r04, gandhi11}) is situated within this torus (R$\sim$5$''$; highlighted in Fig.~\ref{fig:CO+NeII} with a red dashed curve), and has a much steeper position-velocity gradient (Fig.~\ref{fig:x2_models}, right panel). This ionized emission traces gas following the (almost circular) $x_{2}$ orbits \citep{wills00}. The ring is composed of a number of bright clumps; \citet{gandhi11} identified more than 20 discrete sources, 4--5 of which match the location of H\two\ regions identified in the radio by \citet{mcdonald02}, and are consistent with being powered by embedded young super star clusters. 
It seems, therefore (at least at the present moment), that the $x_1$ orbits contain \emph{mostly} neutral/molecular material \citep{wills00} and the
$x_2$ orbits  \emph{mostly} ionized gas, although the presence of compact, self-shielded molecular clumps in the central region cannot be excluded.
Interestingly, M82 was the first barred galaxy to show such a dynamical and spatial distinction between ionized and non-ionized material \citep{wills00}.

In projection, clusters A1, A2 and A3 are located between the eastern edge of the circumnuclear ring and the eastern side of the molecular torus. All three have a velocity consistent with being at one end of the $x_2$ orbits, or possibly the $x_1$ cusp orbits where orbit spraying is occurring \citep{greve02}. We previously  argued that A1 may have formed as a result of intense star-formation in the intersection between the $x_1$- and $x_2$-orbit families \citepalias{westm07c}. Indeed, the Arches cluster in our Galaxy could have also formed in the same circumstances since it is on a transitional trajectory between the Milky Way's $x_1$ and $x_2$ orbits \citep{stolte08}. 
In any case, since we find that A1, A2 and A3 are close in both position and velocity in this region of M82, it is likely that they are also close spatially because the orbital velocities change rapidly moving outwards from the bar (Fig.~\ref{fig:x2_models}).

Thus we have a cold gas reservoir in the form of the molecular torus, a nuclear ring of young star-forming knots, and clusters A1,2,3 located between the two. Did A1,2,3 form in the ring too? And where is the molecular fuel coming from for the SF in the nuclear ring?

\citet{van-de-ven09} performed calculations of the dynamics of a star cluster and gas ring system, including the effects of dynamical friction of the host galaxy, and gas inflows along the bar onto the nuclear ring. Their model assumed a constant and smooth gas mass inflow such that the outer edge of the ring has a surface density enhancement sufficient to form star clusters. Indeed simulations of the Milky-Way nuclear ring show that star formation takes place mostly in the outermost $x_2$ orbits because newly infalling gas collides with the nuclear ring at its outer rim \citep{kim11}. However, \citet{van-de-ven09} note that if the gas inflow is clumpy, or possesses a different vertical scale height or different inclination to the gas ring, the flow may not (only) merge at the outermost radius. Since the nuclear ring is not complete, and composed of multiple clumps, this implies that the gas inflow is not smooth and well-aligned with the ring, but more clumpy and chaotic.

In their simulations, \citet{van-de-ven09} found that star clusters that initially form in a nuclear ring can move radially outwards because of satellite-disk tidal interactions. It is possible, therefore, that A1,2,3 may have formed in the nuclear ring and subsequently moved outwards. \citet{van-de-ven09} predict that the timescale for separation from the ring is of order a few to tens of orbits. The orbital period of the nuclear ring in M82 is $\sim$5~Myr \citep{achtermann95}, and the bar(/torus) $\sim$12~Myr \citep[assuming a bar rotation speed of 140~\kms\ and bar length of 540~pc;][]{wills00}. However the ages of the three clusters are between these two values, meaning they have made $\lesssim$1 orbit. This may not be enough time for the clusters to have migrated so far, thus making the migration scenario unlikely.

The clusters must therefore have formed \textit{in-situ}. In order for them to have done that, there must have been a sufficient supply of molecular gas at their location $>$10~Myr ago. Although, as mentioned above, the $x_2$ orbits and nuclear ring are currently predominantly ionized, there is still some molecular and neutral gas associated with the steeper $x_2$ orbits \citep[and ionized gas with the $x_1$ orbits; Fig.~\ref{fig:x2_models} and][]{wills00}. Given the dynamically active and variable state of the central starburst, it is likely that the state of the gas on the various orbits changes on short timescales. The UV flux is high, and molecular gas will not stay in this phase for long \citep[e.g.][]{lord96}. We therefore propose that clusters A1,2,3 were formed on the $x_2$ orbits in regions of dense molecular gas
and the subsequent negative radiative feedback from the central starburst region has acted to change the state of the gas to what we presently see.

A logical corollary of this is that in the recent past there must also have been a large reservoir of molecular gas in the nuclear ring in order to fuel the formation of the embedded star clusters energising the many H\two\ regions found here \citep{gandhi11}. We might therefore expect the next generation of clusters to form in the dense molecular torus surrounding the $x_2$ orbits.

What will become of these three clusters, A1,2,3? Their similar radial velocities suggest that they may merge. However, they are in a region with a strong velocity shear and thus their fate is not clear without detailed calculations.
\citet{morrison11} recently studied the globular cluster population in M31 and found a number of metal-rich globular clusters on bar orbits.  If they do merge, A1,2,3 could represent the young progenitors of globular clusters in a future bulge.

Another possibility for the future fate of these clusters is that they will dynamically inspiral to form a nuclear star cluster \citep[e.g.][]{tremaine75,antonini13}.  Such nuclear star clusters are seen in $\sim$75\% of galaxies in M82's mass range, regardless of type \citep{boker02,cote06}.  \citet{ebisuzaki01} calculated the dynamical friction timescale for massive clusters to reach the center of M82:\\
\begin{equation*}
t_{\rm fric} = 6 \times 10^{8} \left( \frac{r}{\rm 1 kpc} \right)^2 \left( \frac{v_c}{\rm 100 km/s} \right)^2 \left( \frac{5 \times 10^6 M_\odot}{m} \right)^2
\end{equation*}
where $m$ is the mass of the cluster, $v_c$ is the velocity dispersion, and $r$ is the radius of the cluster.  Assuming the cluster masses and errors given in Table~4, a dispersion similar to the observed circular velocity, $v_c \sim 100\pm30$~km/s, and a true radius  within a factor of $\sqrt{2}$ of the projected radius, we find dynamical friction timescales of less than a billion years for all clusters.  The timescales for A1 range from $2 \times 10^7$ and $5 \times 10^8$ yr, A2 from $4 \times 10^7$ and $9 \times 10^8$ yr, and A3 from $2 \times 10^7$ and $4 \times 10^8$ yr.  Despite this simple formula's exclusion of the internal and external forces that will reduce the cluster mass as it inspirals \citep[e.g.][]{antonini13}, the timescales are short enough that it appears likely that these clusters will reach the center of M82 within the next $\sim$1~Gyr.


\subsection{The environment around A2 and A3}\label{sect:environment}

\subsubsection{Densities and pressures}
One of the surprising findings of \citetalias{smith06} was that cluster A1 has been following a non-standard evolutionary path for a star cluster, since it is still surrounded by a large, pressurised ($P/k = 1$--$2\times10^7$~cm$^{-3}$~K) H\two\ nebula at its age of 6.5~Myr. Standard cluster H\two\ region evolution scenarios predict that the surrounding gas should have been completely cleared away by the cluster winds and radiation by this time. In \citetalias{westm07c} we presented measurements of the surrounding environment to A1 and found similarly high ISM pressures of $P/k= 0.5$--$1.0\times 10^{7}$~cm$^{-3}$~K, leading us to conclude that the high interstellar pressures and strong radiative cooling in the starburst core are acting to stall the cluster winds \citep[see][]{silich07}. 

How do A2 and A3 compare? A2 is located in a region of much lower nebular background emission than A3 (Fig.~\ref{fig:finder}), however its (A2's) H\two\ region is very faint, with peak fluxes only $\sim$2 times over the surrounding diffuse emission. In contrast, the H$\alpha$ emission associated with A1 peaks at a value $\sim$12 times higher than the surroundings, and the emission associated with A3 peaks at $\sim$3 times over that of its surroundings.

\begin{figure}
\centering
\includegraphics[width=0.49\textwidth]{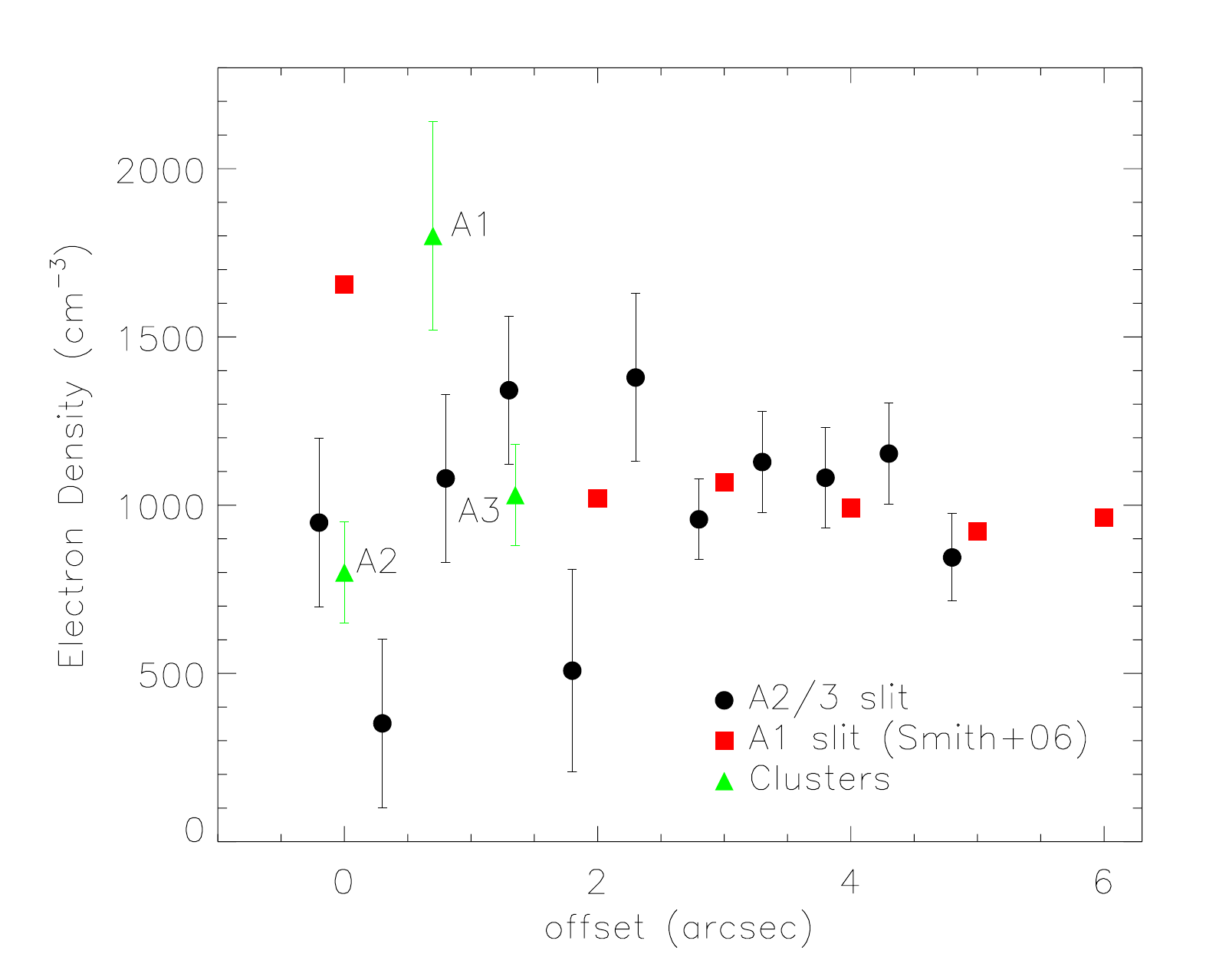}
\caption{Electron densities derived from the [S\two]$\lambda$6717/$\lambda$6731 flux ratio measured in 10 pixel bins along the A2/3 slit, compared to the measurements along the A1 slit from \citetalias{westm07c} and those from the individual cluster spectra.}
\label{fig:ap10elecdens}
\end{figure}

In \citet{westm09b}, we presented spatially resolved IFU spectroscopy of the central $\sim$500~pc of the starburst core, and found that the location of A1 indeed coincides with the electron density ($\equiv$ pressure) peak, and that the densities fall off rapidly to the south and west (where A2 and A3 are located). To investigate further, we measured the electron density from the [S\two]$\lambda$6717/$\lambda$6731 doublet ratio in both of the extracted STIS cluster spectra and found values of $800\pm150$~cm$^{-3}$ for A2 and $1030\pm150$~cm$^{-3}$ for A3. These density differences account well for the variation in H\two\ region brightness described above, since $F({\rm H\alpha}) \propto n_e^2$. We then measured the electron density in the surrounding gas from spectra extracted in 10~pixel bins (in order to get an adequate S/N level) along the slit near A2 and A3. The results are shown in Fig.~\ref{fig:ap10elecdens} together with those measured from the A1 slit \citepalias[located $\sim$parallel, $1\farcs2$ to the north;][]{westm07c}. The average densities of the intra-cluster medium in clump A measured from the two slits are consistent at $\sim$1000--1300~\cmt.


Clusters A2 and A3 are therefore also located in a high density environment which is only a factor of $\lesssim$2 lower than that surrounding A1. We detect compact H\two\ regions surrounding both the clusters with sizes $\sim$2--2.5 times larger than the clusters themselves. Although A2 and A3 are  younger than A1, according to the standard theory of bubble evolution \citep{weaver77} they should still have blown away all their surrounding gas at their ages. If the pressure of the ambient ISM is high enough, it can act to stall the star cluster winds at smaller radii. It is likely that in such a high density environment the transition from the energy- to momentum-dominated regimes (when the cluster wind impacts directly on the shell) occurs very rapidly \citep[see, for example][]{silich13}.

One can estimate then the radius of the stalling shell from the condition that $P_{\rm ram} = P_{\rm ISM}$, where $P_{\rm ram} = L_{\rm mech}/2 \pi v_{\rm inf} R^2$ is the ram pressure in the wind, $P_{\rm ISM}/k = n_{\rm ISM} T_{\rm ISM}$ is the thermal pressure in the surrounding medium, $L_{\rm mech}$ is the mechanical energy input rate, $v_{\rm inf}$ is the wind terminal velocity, and $R$ is the stalling radius. Taking as representative values $v_{\rm inf} = 1000$~\kms\ \citep[e.g.][]{mokiem07}, $T_{\rm ISM}=10^4$~K and $n_{\rm ISM}=1000$~\cmt, we can calculate R for different $L_{\rm mech}$. This yields $R = 24.1$~pc for $L_{\rm mech} = 10^{40}$~erg~s$^{-1}$, $R = 7.6$~pc for $L_{\rm mech} = 10^{39}$~erg~s$^{-1}$, and R = 5.4~pc for $L_{\rm mech} = 3\times10^{38}$~erg~s$^{-1}$. The mechanical luminosity of a $5 \times 10^5$~\Msol\ cluster with a standard Kroupa IMF and solar metallicity at an age of 4--5~Myr is $L_{\rm mech} = (1-2) \times 10^{40}$~erg~s$^{-1}$ \citep{leitherer99}. We therefore conclude that, like in the case of A1 \citep{silich07}, the winds from A2 and A3 have also been stalled due to the high ambient pressure in this region of M82 and that a significant fraction of their mechanical power must be lost inside the clusters due to strong radiative cooling. This retention of material that would otherwise be expelled could play a role in providing the material for a second star-formation episode, as seen in GC self-enrichment scenarios \citep{conroy11, gratton12}.

These results imply that either lower mass clusters with negligible cooling rates and field stars are more significant as a driving agent for the M82 superwind, or that only the clusters located outside of the high pressure central regions \citepalias{smith06} and the interaction between mechanical and radiative feedback processes acting on different temporal and spatial scales \citep{hopkins12} are important for driving galactic winds.

\section{Summary}

In \citet[][\citetalias{smith06}]{smith06}, we examined a bright, isolated, super star cluster (SSC) in region A of the M82 starburst designated M82-A1. Our \textit{HST}/STIS spectroscopy and supporting imaging allowed us to determine the age ($6.4\pm0.5$~Myr) and mass (7--$13\times10^5$~\Msol) of the cluster. We found this cluster to be surrounded by a compact H\two\ region at high pressure, in pressure equilibrium with the surrounding ISM \citep[][\citetalias{westm07c}]{westm07c}. This led us to conclude that these high intra-cluster gas pressures have caused the A1 cluster wind to stall \citep[or stagnate;][]{silich07}.

Here we present a study of two neighbouring clusters, which we designate A2 and A3, using new \textit{HST}/STIS optical spectroscopy. We also present Gemini/NIFS near-IR IFU spectroscopy of all three clusters and their surrounding medium.

From Starburst99 \citep{leitherer99} evolutionary synthesis model fitting of the optical STIS spectra, we find a degeneracy in the age solution for clusters A2 and A3 where a range from 3.5--10~Myr (and reddenings of $E(B-V) =1.2$--1.9~mag) is equally acceptable, whereas A1 has a clear solution at 6--7~Myr (with $E(B-V)=1.3$~mag). However, the extracted H-band spectra of the three clusters from our NIFS observations show that while A1 exhibits strong CO absorption features arising from RSGs, they are weak or absent in both A2 and A3. Since the expected age for the appearance of the first RSG in a star cluster that fully samples the IMF is $5.7\pm0.8$~Myr \citep{gazak13}, this allowed us to break the degeneracy in our optically-derived ages. The absence or weakness of RSG features places A2 and A3 at ages of $4.5\pm1.0$~Myr. Thus the three clusters straddle the important epoch of the onset of the RSG phase at $\sim$6~Myr. This appearance of RSGs represents an extremely accurate age-dating tool, which, if properly calibrated, will be very important in the era of high sensitivity and/or wide field near-IR instrumentation with e.g.\ VLT/KMOS or \textit{JWST}.

\textit{HST}/ACS-HRC photometry of the three clusters allowed us to derive estimates of their photometric masses using the GALEV models \citep{kotulla09}. We find that the masses of the three clusters are surprisingly similar, in the range 4--$7\times10^5$~\Msol. Those for M82-A1 are in good agreement with the results derived in \citetalias{smith06}. We measured sizes of the three clusters using the \textit{HST}/ACS-HRC imaging, and found A1,2,3 to have $R_{\rm eff}=159$~mas ($\sim$2.8~pc), 104~mas ($\sim$1.8~pc), and 59~mas ($\sim$1.0~pc), respectively. A2 and A3 are therefore surprisingly compact for their mass. Finally, we measured the radial velocities of the three clusters using the nebular emission lines from their compact H\two\ regions and found heliocentric radial velocities of $320\pm2$~\kms, $330\pm6$~\kms, and $336\pm5$~\kms\ for A1,2,3, respectively, placing the three clusters at the eastern end of the $x_2$ orbits.

An investigation of the immediate environments of A2 and A3 show that they are located in a high density environment which is only a factor of $\lesssim$2 lower than that surrounding A1. We detect compact H\two\ regions surrounding both the clusters with sizes $\sim$2--2.5 times larger than the clusters themselves. At their masses and ages, their predicted mechanical luminosities should have blown bubbles of much larger size, meaning that a significant fraction of this input energy has been lost inside the cluster due to strong radiative cooling and that, like in the case of A1, the winds from A2 and A3 have been stalled due to the high ambient pressure of the ISM in this region of M82. 

We discuss possible formation scenarios of these three clusters, given all that we know about the kinematics and gas distribution within the central starburst zone \citep[e.g.][]{larkin94, achtermann95, lord96, weis01, r-r04, westm07c, fuente08, gandhi11}. We consider two possibilities: that the clusters formed within the (currently ionized) circumnuclear ring and subsequently migrated outwards, or that they formed \textit{in-situ}. Since the time needed to migrate this far away from the ring almost certainly exceeds the ages of the clusters \citep{van-de-ven09}, we propose that they were formed \textit{in-situ} on the outer $x_2$ orbits of the well-known bar in previously existing regions of dense molecular gas. Presently, the $x_2$ orbits contain mostly ionized gas, while it is the $x_1$ orbits that contain most of the neutral/molecular material \citep{wills00}. Therefore, we suggest that the subsequent negative radiative feedback from the central starburst region has acted to change the state of the gas to
what we presently see. This is consistent with the dynamically active state of the central starburst, where the state of the gas on the various orbits is likely to change on short timescales. 

The similar radial velocities of A1, A2 and A3 and their small projected separation of only $1\farcs5$ ($\sim$25~pc) suggest that the three clusters may merge in the near future, although this may be prevented by  strong velocity shears in this region of M82.
The very similar properties that we have derived for the three clusters in terms of ages, masses, and velocities suggest that A1, A2 and A3 are indeed a super star cluster triplet.

\section*{Acknowledgments}
MSW would like to thank Charles Proffitt for his help in reducing the STIS observations. We thank the referee for very useful comments, which improved the paper. The research leading to these results has received funding from the European Community's Seventh Framework Programme (/FP7/2007-2013/) under grant agreement No 229517. JSG and RWO gratefully acknowledge partial support of this research through program GO-11641 by the Space Telescope Science Institute, which is operated by the Associated Universities for Research in Astronomy, Inc., under NASA contract NAS 5-26555. 
This work is partly supported by CONACyT (Mexico) research grants CB-2010-01-155142-G3 (PI:YDM) and CB-2011-01-167281-F3 (PI:DRG).
SS gratefully acknowledges support through CONACYT - Mexico, research grant 131913.

\clearpage
\bibliographystyle{apj}
\bibliography{references}

\end{document}